\documentstyle[11pt,newpasp,twoside,epsf]{article}
\def\kms {\ifmmode{{\rm ~km\,s}^{-1}}\else{~km~s$^{-1}$}\fi}

\def\edcomment#1{\iffalse\marginpar{\raggedright\sl#1\/19}\else\relax\fi}
\reversemarginpar

\begin{document}
\title{Extragalactic H2O Masers}
\author{Lincoln J. Greenhill}
\affil{Harvard-Smithsonian Center for Astrophysics, \\
60 Garden St., M.S. 42, \\
Cambridge, MA 02138 USA}

\begin{abstract}
Study of extragalactic H$_2$O masers has progressed significantly in the 25 years since their discovery. 
Existing in star forming regions and in the accretion disks  supermassive black holes, they are familiar
and unfamiliar at the same time. A review of how our understanding has grown, up to the present day, is
followed by comments on future prospects. 

\end{abstract}

\section{Historical Perspective}

Galactic water maser emission in regions of star formation had been known for about a decade (Cheung et al.
1969) when Churchwell et al. (1977) discovered the first extragalactic source, in M\,33 (Table\,1). 
Several unsuccessful surveys of nearby galaxies had been conducted before the Churchwell et al. work
(Dickinson \& Chaisson 1971; Sullivan 1973;  Andrew et al. 1975), concentrating on nuclei and fields around
H\,II regions in the stellar disks. Ultimately, the location and origin of the maser in M\,33 was not
surprising, as it lay in IC\,133, a prominent star forming region in the outer stellar disk.  

\begin{table}[ht]
\textwidth 4.0in
\caption{Historical Perspective}
\begin{tabular}{llc}

\tableline
Date\tablenotemark{(a)} & {\hfill Event \hfill} & Reference \\
\tableline

11/76	  & First extragalactic H$_2$O maser, in M\,33     & 1       \\
09/77	  & First nuclear maser, in NGC\,4945\tablenotemark{(b)} & 2 \\
11/82   &	OVRO survey of late-type galaxies              & 3       \\  
09/83   & First VLA \& VLBI detections of compact structure& 4     \\  
10/84   &                                                &  5      \\
01/92   & High-velocity lines discovered, in NGC\,4258   & 6       \\
        & First wide bandwidth survey of 11 AGN          & 7       \\   
03/93   & Narrow bandwidth surveys of 100s of AGN        & 8, 9    \\  
04/94   & Discovery that NGC\,4258 masers trace sub-pc accretion disk & 10, 11 \\ 
04/95	  & Detection of $6000 L_\odot$ (isotropic) maser at $\sim100$ Mpc & 12     \\ 
08/95    & First associations of masers with radio jets   & 13      \\ 
11/95   &                                                & 14      \\
06/97   & Discovery of masers in an AGN wind             & 15      \\ 
Today   & Continuation of wide bandwidth surveys of AGN  & 9,16    \\ 
\tableline
\tableline
\tablenotetext{(a)}{Date of related observations (in $mm/yy$ format) in all but one case. The report of
weak jet-related maser emission in NGC\,1068 by Gallimore et al. (1996) relied upon archival data obtained
with the VLA by M. Claussen in 1983 and 1987. The paper was submitted in 1995 August.  Data supporting the
report of more intense emission excited by the jet in the radio galaxy NGC\,1052  was obtained with the
VLBA in 1995, again by M. Claussen. }

\tablenotetext{(b)}{L\'epine \& Dos Santos (1977) report an earlier detection of $5\pm1$ Jy emission toward
the nucleus of NGC\,253 at an LSR velocity of 253\,km\,s$^{-1}$.  However, later observations by
Batchelor, Jauncey, \& Whiteoak (1982), Nakai \& Kasuga (1988), and Ho et al. (1987) only found emission on
the order of tenths of Jy and no emission within 100\,km\,s$^{-1}$ of the velocity reported by L\'epine \&
Dos Santos. }

\tablenotetext{}
{Citations: (1) Churchwell et al. (1977); (2) Dos Santos \& L\'epine (1979); (3) Claussen, Heiligman, \&
Lo (1984);  (4) Claussen \& Lo (1986); (5) Claussen et al. (1988);  (6) Nakai, Inoue, \& Miyoshi (1993);
(7) Nakai et al. (1995); (8) Braatz et al. (1996); (9) Greenhill et al. (2002);  (10) Greenhill et al.
(1994); (11) Miyoshi et al (1995); (12) Koekemoer et al. (1995);  (13) Gallimore et al. (1996), (14)
Claussen et al. (1998); (15) Greenhill et al. (2000);  (16) Braatz et al., Henkel et al., Nakai et al., and
others, unpublished. }

\end{tabular}
\end{table}

\begin{table}[ht]
\textwidth 4.0in
\caption{Origin of Extragalactic H$_2$O Masers: Evolution of Thought}
\begin{tabular}{lllc}
\tableline
Year      & Impetus                            & Realization        & Reference \\
\tableline
1976	  & Masers in the disk of M\,33           &	Association with star formation        & 1    \\ 
1977   & Masers in a star-forming nucleus      & High apparent luminosities possible    & 2    \\ 
1983	  & Compact ($\la 10$ pc) structure       & Axially thick, circumnuclear disks     & 3    \\
       &                                       & Probably not star-formation related    & ...  \\
1992+  &	Surveys                               & Certain association with AGN           & 4    \\ 
1994   & Structure of the NGC\,4258 maser      & Origin in thin accretion disks	        & 5, 6 \\
1995 	 & Locations of NGC\,1068/1052 masers    & Origin in jet-excited material         & 7, 8 \\ 
1997   & Distribution of masers in Circinus    & Origin in nuclear winds                & 9    \\

\tableline
\tableline
\tablenotetext{}
{Citations: (1) Churchwell et al. (1977); (2) Dos Santos \& L\'epine (1979); (3) Claussen \& Lo
(1986); (4) Braatz et al. (1996); (5) Greenhill et al. (1994, 1995b); (6) Miyoshi et al (1995); 
(7) Gallimore et al. (1996); (8) Claussen et al. (1998); (9) Greenhill et al. 2000a. }
\end{tabular}
\end{table}

The discovery of the first maser source in a galactic nucleus led to speculation that it was a beacon of
particularly intense star formation (Table\,2), of the type observed toward starbursts (Dos Santos \&
L\'epine 1979; see also Moorwood \& Glass 1983).  This first ``nuclear maser'' lay in NGC\,4945 (Table\,1),
which is one of the most luminous nearby far-infrared (FIR) galaxies, and which hosts a massive
circumnuclear star-forming ring or disk of gas on 100 kpc scales (Moorwood et al. 1996).  Because star
formation rate is correlated with FIR emission, the subsequent discovery of more masers in FIR-bright
galaxies was apparently easy to understand.

Interferometric observations of the structure of nuclear masers demonstrated that they have
more compact structure that was expected for emission tied to distributed star formation in a
starburst.  Claussen \& Lo (1986) used the VLA to demonstrate that the masers in NGC\,1068
and NGC\,4258 were compact on scales of less than several parsecs.  Although nuclear starbursts can also be
compact, the observed compactness was in fact the first clue pointing to a truly extraordinary phenomenon.
In light of the Antonucci \& Miller (1985) model for AGN structure, Claussen \& Lo conjectured that the
emission was tied to an interaction between nuclear outflow and gas in circumnuclear tori.  

The years 1986 to 1992 marked a drought during which no H$_2$O masers were discovered in new galaxies
(though a second weak maser was found in the nearby dwarf irregular galaxy IC\,10 (Becker et al. 1993)). 
Up to this time, searches had emphasized FIR-bright and nearby star-forming galaxies, and nine of the eleven
then known extragalactic H$_2$O masers were associated with the 83 known IRAS galaxies with $100\mu$m flux
density $>50$ Jy.   However, this selection criteria was misdirected. In retrospect, discovery of these nine
masers may have depended more on proximity than  on a (hoped for) direct physical relationship between maser
emission (which we now know often arises in parsec scale structures) and IRAS far-infrared emission (which
also originates on scales that are orders of magnitude larger).  Among the larger number of H$_2$O masers
known today, there is no apparent correlation.  Galaxies with similar IRAS
$100\mu$m flux densities can have peak maser flux densities that differ by over an order of magnitude, and
visa versa. 

In the early 1990s there were two watershed discoveries in the study of extragalactic H$_2$O masers and
perhaps AGN.  

First, Braatz et al. (1996) discovered 11 new H$_2$O masers by targeting AGN, achieving sensitivities
($3\sigma$) on the order of 0.1 Jy, for galaxies mostly closer than 100 Mpc.   This survey established a
strong link between H$_2$O maser emission and AGN, specifically Seyfert 2 objects and Low Ionization
Nuclear Emission Regions (LINERs).  Though other surveys also concentrated on AGN (e.g., Nakai et al. 1995;
Greenhill et al. 1995; Greenhill et al. 2002), they did not enjoy as large a sample of nearby galaxies and
as uniformly high sensitivity.  

Second, Nakai, Inoue, \& Miyoshi (1993) made the serendipitous discovery of high-velocity maser lines
symmetrically bracketing the already known emission near the galactic systemic velocity  in NGC\,4258. 
Their interpretation included three possible models: Raman scattering, outflow, or
rotation, the latter being of particular interest given the earlier work of Claussen \& Lo (1986).  Very
Long Baseline interferometric (VLBI) observation of the then known maser emission near the systemic
velocity had been conducted in 1984 (Claussen et al. 1988), and the preliminary reduction revealed little
specific structure.  However, reanalysis of the data, interpreted in the context of the newly discovered
high-velocity lines (Greenhill et al. 1994, 1995b) demonstrated that the emission probably arose in a
sub-parsec diameter disk bound by a very massive and compact object.  

Support for the disk model arrived from the {\it independent}  measurement of the line-of-sight
(centripetal) acceleration of the maser-emitting gas close to the systemic velocity.  Observation of a
drift in the line-of-sight velocity of one spectral feature by $\sim 10$\,km\,s$^{-1}$\,yr$^{-1}$ (Haschick
\& Baan 1990) was not widely accepted because in general the NGC\,4258 spectrum is a blend of many time
variable features, and specifically, it was not possible to prove that a single feature had been tracked.
However, analysis of archival data (1984-1986) obtained at Effelsberg (Greenhill et al. 1995a) and further
observations by Haschick, Baan, \& Peng (1994) demonstrated that on the order of 10 features distributed
throughout the spectrum drifted in velocity at about the same rate, which solidified the interpretation of
detectable acceleration.  The magnitude of the drift matched the predicted centripetal acceleration given
the inferred disk rotation speed and radius, for an assumed distance of $\sim 7$ Mpc (Greenhill et al.
1994; Watson \& Wallin 1994).  

Observation of the NGC\,4258 H$_2$O maser in mid-1994 with the then newly commissioned Very Long
Baseline Array (VLBA), provided conclusive evidence in support of the disk model (Miyoshi et al. 1995).  The
VLBA observed  both the systemic and high-velocity emission simultaneously with high angular resolution for
the first time.   The resulting model parameters were quite surprising: the disk obeyed a Keplerian rotation
curve with deviations of $<1\%$; the molecular gas lay as close as 0.14 pc from a $3.9\times10^7$ M$_\odot$
central object; the disk was thin, with a ratio of height to radius of $\ll 1\%$; and the disk was
misaligned and  counter-rotating with respect to the galactic disk. (N.B. Parameter values reflect updates
by Herrnstein et al. (1999) following the measruement of a precision geometric distance to
NGC\,4258, based on maser proper motions and accelerations.  The distance, $7.2\pm0.5$ Mpc, is the most
precise yet measured for galaxy that is independent of the calibration of any Standard Candle.)

NGC\,4258 defined a paradigm for the interpretation of spectra and VLBI images for other extragalactic
H$_2$O maser, especially sources with high-velocity emission (e.g., NGC\,1068).  However, that paradigm has
not been appropriate in all cases.  First, the masers in NGC\,1052 (Claussen et al. 1998) and Mrk\,348 (Peck
et al. 2001) exhibited distinctive, singular, broad emission features (50-100\,km\,s$^{-1}$), which are
unlike the complexes of narrow lines associated with masers in accretion disks.  VLBI observations have
demonstrated conclusively that these maser source lie offset from their respective central engines
entirely, and in close association with radio jets.  In addition, VLA observations of the NGC\,1068 maser
have detected a site of H$_2$O maser emission in the vicinity of a jet-cloud collision $\sim 20$ pc from
the central engine (Gallimore et al. 1996).   These sources  may be stimulated directly by jet-activity,
perhaps as a result of entrainment or shock heating of ambient material. Two other sources may be similar,
IRAS\,F22265-1826 (Greenhill et al., in prep) and IRAS\,F01063-8034 (Greenhill et al. 2002), for which
VLBI data is inconclusive or not yet available. Second, the maser source in the Circinus galaxy exhibits 
two loci of maser emission, only one of which corresponds to an accretion disk.  The second locus appears to
trace directly structure in a wide-angle wind that originates $\la 0.1$ pc from the central engine.  The
boundaries of the wind are set by the (warped) accretion disk and correspond to the boundaries of the
ionization cone that has been detected on scales $>100$ pc (Veilleux \& Bland-Hawthorn 1997).  When nuclear
maser emission was  first recognized to lie in accretion disks and in close proximity to super massive
black holes, it was a surprise.  The association of maser emission with jets and with the innermost reaches 
of an outflowing nuclear wind are surprises nearly as significant.

\section{Today}

Over 1000 galaxies have been studied in the hope of detecting H$_2$O maser emission. Table\,3 lists seven
galaxies for which H$_2$O maser emission is believed to be excited by star formation processes in a total
of 19 regions. Except for sources in the Magellanic Clouds and one source in IC\,10, which has been
observed to reach 120 Jy (Baan \& Haschick 1994), the known masers in extragalactic star forming regions
are weak and hence, challenging to study.  VLBI images of the strongest two source in the northern
hemisphere, IC\,10SE and IC\,133, have been published.  The distribution of masers in IC\,10SE (Argon et
al. 1994) subtends $\sim 0.1$ pc.  Emission in the IC\,133 region comprises two regions that each subtend
$\sim 0.1$ pc and are separated by $\sim 0.3$ pc (Greenhill et al. 1993). All three centers of maser
activity are coincident with thermal radio continuum sources that are probably compact HII regions.  Proper
motions on the order of 1 to 10 $\mu$as\,yr$^{-1}$ have been measured for the IC\,133 H$_2$O masers, using
two epochs of data (Greenhill et al. 1993). The motions are suggestive of a poorly collimated bipolar
outflow, as are observed in galactic high mass star formation.  Unfortunately, no published studies
exist of the structure and evolution of H$_2$O masers seen toward starbursts (i.e. M\,82 and NGC\,253),
wherein star formation takes place in a rather more exotic environment.

\begin{table}[ht]
\textwidth 4.0in
\caption{Extragalactic H$_2$O Masers \\ Associated with Star Formation}
\begin{center}
\begin{tabular}{llllc}
\tableline
Galaxy & Distance & F$_\nu$\tablenotemark{(a)} & No.  & Ref.\\ 
& (Mpc) & (Jy) & & \\
\tableline

LMC  	    & 0.05	& 3-20	    & 6  & 1, 2, 3, 4 \\
SMC  	    & 0.05	&  4-7    	& 2  & ...        \\
M\,33	    & 0.7  &    1    	& 5  & 5, 6, 7    \\
IC\,10    &	1.3	 & 0.2-120  &	2  & 8, 9       \\
NGC\,253	 & 2.5	 &  0.1    	& 1  & 10, 11?    \\
M\,82	    & 3.3	 &  0.2	    & 1  & 12         \\
IC\,342	  & 3.9	 & 0.2--0.4 &	2  & 13         \\
M\,101    &	2.5	 & marginal & 1  & 7          \\
NGC\,2403 &	3.2	 & marginal & 1  & 7          \\
NGC\,2366 &	3.2	 & marginal & 2  & 7          \\
NGC\,672  & 8    & marginal & 2  & 7          \\

\tableline
\tableline
\tablenotetext{(a)}{Characteristic peak flux density or that at time of discovery.}
\tablenotetext{}{Citations: (1) Scalise \& Braz (1981); (2) Scalise \& Braz (1982); 
(3) Whiteoak et al. (1983); (4) Whiteoak \& Gardner (1986); (5) Churchwell et al. (1977);  
(6) Huchtmeier et al. (1978); (7) Huchtmeier et al. (1988); (8) Henkel, Wouterloot, \& Bally (1986); 
(9) Becker et al. (1993); (10) Ho et al. (1987); (11) L\'epine \& Dos Santos (1977); (12) Claussen et al.
(1984);  (13) Huchtmeier et al. (1978). }

\end{tabular}
\end{center}
\end{table}

Water maser emission has been detected and confirmed in 22 AGN (Table\,4). Surveys have achieved a 5-10\%
detection rate for samples of nearby Seyfert\,2 galaxies and LINERs (e.g., Braatz et al. 1996) that
declines with distance, probably because of limited instrument sensitivity. At present IRAS\,F22265-1826
and NGC\,6240 contain the most distant known masers, at $\sim 100$ Mpc. A survey of more distant
Fanaroff-Riley type 1 radio galaxies (Henkel et al. 1998) and radio-bright quasars that are part of the
VLBA calibrator survey (Herrnstein, Beasley, \& Greenhill, unpublished) have not detected any maser
emission. (See also Matsakis et al. 1982.) 

The classification of masers depends on spectroscopic signatures and where available, on structural
signatures (i.e., angular distributions of emission mapped with VLBI).  For purposes of taxonomy,  H$_2$O
maser sources may be grouped into four classes: disk origin, jet origin, peculiar, and unknown. 

Of the known maser sources, three exhibit the spectroscopic and angular (i.e., VLBI) signatures of accretion
disks: NGC\,4258 (Miyoshi et al. 1995), NGC\,1068 (Greenhill \& Gwinn 1997), and Circinus (Greenhill et al.
2000a). Two other sources, for which imaging is incomplete or absent, exhibit (only) the spectroscopic
signatures of disks with rotation speeds on the order of a few hundred km\,s$^{-1}$: NGC\,5793 (Hagiwara et
al. 1997), and one as yet unpublished (Braatz, private communication).  One source, NGC\,2639, though
without detectable high-velocity emission has displayed a drift in Doppler velocity that is
reminiscent of the centripetal acceleration among masers in NGC\,4258 (Wilson et al. 1996). 

The process by which maser emission is stimulated by jet activity in AGN is not understood, in particular
the reason for the observed unusually broad line profiles. Four sources display these line profiles:
NGC\,1052 (Braatz et al. 1996), Mrk\,348 (Falcke et al. 2000a), IRAS\,F22265-1826 (Koekemoer et al. 1995),
and IRAS\,01063-8034 (Greenhill et al. 2002).  As previously mentioned, the first two sources demonstrate
certain association with jets (i.e. the maser emission is offset from the presumed position of the
respective central engines, and it lies along the line of sight to the jet).  The second two sources are
classified (here) as jet-induced because of circumstantial evidence.  They exhibit broad line profiles but
little more evidence is available. VLBI imaging of IRAS\,F22265-1826 (Greenhill et al., unpublished) has
not detected radio emission from the known optical jet in that source (Falcke et al. 2000b), and
IRAS\,F01063-8034 has been too weak to observe with VLBI, though the centroid velocity of the emission has
been observed to change abruptly, a behavior otherwise unique to the NGC\,1052 maser.

The class of peculiar masers includes three prominent sources that have suggestive spectroscopic
characteristics (e.g., emission close to the systemic velocity but perhaps only one-sided high-velocity
emission): NGC\,3079 (Trotter et al. 1998; Sawada-Satoh et al. 2000), NGC\,4945 (Greenhill et al.,
unpublished, and NGC\,1386 (Braatz et al. unpublished).  Though these sources have highly elongated angular
distributions of masers, the distributions are broadened and display complex velocity structure.  It is
possible that the accretion disks of these galaxies are intrinsically thicker due to properties of the
AGN.  Alternatively, masers may occupy clumpy, fragmented material that lies beyond the outer edges of thin
(``maser-dark'') accretion disks.

The remaining 11 nuclear maser sources not discussed so far fall in the ``unknown'' category.  Most of these
sources have nondescript spectra (e.g., a single line offset from the systemic velocity of the host
galaxy), and many are too weak to be targets for VLBI imaging, excepting all but the most heroic efforts.
However, because masers are time-variable, monitoring of known sources is important because line
intensities can increase on time scales of weeks to months, and new spectral features can appear that
would make possible ready classification.

\begin{table}[ht]
\caption{Established Cases of H$_2$O Masers in AGN\tablenotemark{(a)}}
\begin{tabular}{lllllll}
\tableline
Galaxy & Distance\tablenotemark{(b)} & F$_\nu$\tablenotemark{(c)} &
Galaxy & Distance\tablenotemark{(b)} & F$_\nu$\tablenotemark{(c)} \\
& (Mpc) & (Jy) &  & (Mpc) & (Jy)  \\
\tableline

NGC\,4945	& 3.7	& 4		 & IC\,2560          &  38 & 0.4 \\
Circinus	 & 4	  & 4		 &	NGC\,2639	        &  44 & 0.1	\\	
NGC\,4258	& 7.3	& 4		 &	NGC\,5793	        &  50	& 0.4	\\
M\,51	    & 9.6	& 0.2 & ESO\,103-G35	     &  53 & 0.7	\\
NGC\,3079	& 16	 & 6		 &	Mrk\,1210	        &  54	& 0.2 \\
NGC\,1068	& 16	 & 0.6	&	ESO\,013-G12      &  57 & 0.2 \\
NGC\,1386	& 12	 & 0.9	&	Mrk\,348	         &  63	& 0.04\\
NGC\,1052	& 20	& 0.3	 & Mrk\,1	           &  65 & 0.1 \\
NGC\,5506	& 24	& 0.6	 & IC\,1481	         &  83	& 0.4 \\
NGC\,5347	& 32	& 0.1	 & NGC\,6240         &  98 & 0.03\\
NGC\,3735	& 36	& 0.2	 &	IRAS\,F22265-1826	& 100	& 0.3	\\

\tableline
\tableline
\tablenotetext{(a)}{Sources of maser emission whose confirmation by more than
one observations has been reported in the literature. Marginal detections for
NGC\, 3227 (Huchtmeier et al. 1988) and NGC\,6946 (Claussen et al. 1984) remain unconfirmed.}

\tablenotetext{(b)}{Distances estimated directly from optical heliocentric velocity, assuming H$_\circ$=75
\kms\,Mpc$^{-1}$.}

\tablenotetext{(c)}{Characteristic peak flux density or flux density at time of discovery.}

\tablenotetext{}{References -- NGC\,4945: Dos Santos \& L\'epine (1979) --  Circinus galaxy: Gardner \&
Whiteoak (1982) -- NGC\,1068, NGC\,4258: Claussen et al. (1984) -- NGC\,6240: Claussen et al.
(1984), Haigwara, Diamond, \& Miyoshi (2001) -- NGC\,3079: Henkel et al. (1984), Haschick \& Baan (1985) --
Mrk\,1, Mrk\,1210, NGC\,1052, NGC\,1386,  NGC\,2639, NGC\,5506, NGC\,5347, NGC\,5793, ESO\,103-\,G\,35,
IC\,1481, IC\,2560: Braatz et al. 1996 -- IRAS\,F22265-1826: Koekemoer et al. 1995 -- M\,51: Ho et al.
(1987) -- NGC\,3735: Greenhill et al. 1997 -- Mrk\,348: Falcke et al. 2000a -- IRAS\,F01063-8034: Greenhill
et al. 2002. 
}

\end{tabular}
\end{table}

\section{Tomorrow}

Extragalactic water maser emission, especially when studied with VLBI techniques, allows the radio
astronomer to establish windows into regions around high-mass protostars and AGN central engines that are
otherwise (1) obscured by considerable visual and infrared extinction and (2) too small to resolve without
overwhelmingly large infrared apertures.  Relatively few extragalactic H$_2$O masers are known today. 
However, those masers that are known make possible exquisitely detailed studies of specific cases, against
which theories may be tested and generalizations constructed.

Extragalactic H$_2$O masers that are excited by star formation are understudied with respect to their 
counterparts in AGN, but they are no less valuable as astronomical probes.  For example, differential
astrometric measurement of maser positions permit the direct observation of galactic rotation in spiral
galaxies,  the estimation of peculiar motions, and the deduction of precision geometric distances.   With
ground-based instruments, these measurements are feasible for targets in the Local Group, such as M\,33. 
However, the most immediately important target may be the LMC, for which
uncertainty in distance is a dominant source of systematic error in the extragalactic distance scale, as
now calibrated by observations of Cepheid variable stars (Freedman et al. 2001).  

The scientifically profitable case of NGC\,4258 remains the best motivation for continued study of
extragalactic H$_2$O masers in AGN.  Among known sources, the uniqueness of the rapidly rotating (and
therefore) thin, quiescent accretion disk in NGC\,4258 begs the question, are other similar sources
detectable?  A great many probably exist though detection depends on many factors, among which are (1) the
anisotropic beaming of maser radiation (e.g., the NGC\,4258 maser is only visible to
$<10\%$ of the Universe), (2) warps in accretion disks, (3) chance alignments among (amplifying) maser
regions and background nonthermal continuum sources, and (4) the range of luminosities among masers.  
Nonetheless, pursuit of high-sensitivity, broadband surveys is necessary to answer the question.  
One NGC\,4258-type maser out of more than $>1000$ candidates searched is strongly suggestive of
an answer.  However, many past surveys relied upon spectrometers with bandwidths corresponding to
$<800$\,km\,s$^{-1}$ (at a rest frequency of 22.2 GHz) and concentrated on emission close to the systemic
velocity of each AGN observed.  Of all masers with known high-velocity emission, only {\it one} displays
its strongest emission close to the systemic velocity of the host galaxy (due to chance alignment with a
background source).  Hence, though successful, much past survey work may have been systematically biased
against discovery of the type of source that is of the greatest interest.  Recent surveys have begun to
address this issue through the use of new instrumentation (see Braatz, this volume).  

Study of extragalactic H$_2$O masers has progressed significantly in the 25 years since their discovery. 
They are familiar and unfamiliar at the same time, existing in (common) star forming regions and in the
(exotic) accretion disks of supermassive black holes.  VLBI observations, using intercontinental baselines,
are critical to the use of these masers as probes of other galaxies and the local Universe. The VLBA and
affiliated antennas is the bedrock of maser studies. The Australian Telescope Long Baseline Array
has been a capable counterpart for southern sources however, development of  a follow-on, VLBA-style,
intercontinental array in the southern hemisphere would be critical to many studies, including measurement
of the LMC distance.  Such an array could form a precursor to the Square Kilometer Array, which is unlikely
to be built until at least the middle of the next decade.  A next-generation space VLBI mission would,
equipped with a broad-band $\lambda 1.3$ cm receiver package and the necessary instrumentation to make
possible precision spectral-line calibration, would perhaps be even more important than a southern VLBA,
because it would permit extended study of many sources that have already been well observed and the further
testing of astronomical theory.

\acknowledgements

I would like to acknowledge my co-authors whose names do not appear in the list of references:  R. Becker,
R. S. Booth, M. J. Claussen, S. P. Ellingsen, R. G. Gough, C. Henkel, J. R.  Herrnstein, D. L. Jauncey, D.
R. Jiang, K.-.Y Lo, P. M. McCulloch, P. J. McGregor, J. M. Moran, R. P. Norris,  C. J. Phillips, D. P.
Rayner, M. J. Reid, J. E. Reynolds, M. W. Sinclair, A. K. Tzioumis, T. L. Wilson, and J. G. A. Wouterloot.

\end{document}